\documentstyle[12pt, epsf]{article} 
\normalbaselines

\newcommand{\be}{\begin{equation}}
\newcommand{\ee}{\end{equation}}

\def\be{\begin{equation}}
\def\ee{\end{equation}}
\def\bea{\begin{eqnarray}}
\def\eea{\end{eqnarray}}

\begin{document}

\begin{center}

{\Large Quantum theory as efficient representation of probabilistic information}

\vspace{6mm}

Johann Summhammer

Vienna University of Technology

Atominstitut

Stadionallee 2 

1020 Vienna, Austria

E-mail: summhammer@ati.ac.at
\end{center}

{\small
Quantum experiments yield random data. We show that the most efficient way to store this empirical
information by a finite number of bits is by means of the vector of square roots of observed 
relative frequencies.
This vector has the unique property that its dispersion becomes {\it invariant}
of the underlying probabilities, and therefore {\it invariant of the physical parameters}.
This also extends to the complex square roots, and it remains true under a unitary transformation. 
This reveals quantum theory as a theory for making predictions which are as accurate as the 
input information, {\it without} any statistical loss.
Our analysis also suggests that from the point of view of information a slightly more accurate 
theory than quantum theory should be possible.
}

\section{Introduction}

There have been several attempts to find an explanation for quantum theory by looking at it as a
theory of information. For instance, Wheeler's work is based on statistical distinguishability \cite{Wheeler},
von Weizs\"acker's {\em ur}-hypothesis starts with empirical yes-no decisions \cite{Weizsacker}, 
Bohr and Ulfbeck emphasize symmetry \cite{ABohr}, Brukner and Zeilinger define an elementary system as answering
only yes or no to any question \cite{Caslav} (see also the essay \cite{Zeilinger2}), and Hardy introduces five
axioms containing no traditional physical concepts \cite{Hardy}. Hardy also cites older axiomatic approaches. 
Grinbaum basis his
derivation of the quantum formalism explicitly on information \cite{Grinbaum}. Luo makes use of Fisher 
information to find Malus' law \cite{Luo}. Mehrafarin derives interference from empirical input 
information \cite{Mehrafarin}. Recently, Aerts exposed quantum theory as a theory of optimal observation 
and emphasized a similarity to the theory of signal analysis \cite{SvenAerts}. Grangier gives a compact
derivation of quantum theory based on the discreteness of the empirical information in quantum experiments
(e.g. \cite{Grangier} and references therein), which is not unlike the work of Land\'e \cite{Lande}.
But also approaches based on structures inherent in probability theory, 
like the one of Caves et al. (e.g. \cite{Caves}) or of Saunders \cite{Saunders}, can be seen as putting 
primacy on the concept of information, since probability is a way of quantifiying information. 

The present paper takes motivation from these works and focusses on a point which does not seem to 
have been touched yet \cite{Summhammer}: The raw data of quantum experiments, as generic probabilistic 
experiments, are random numbers. One may then ask on a purely informational level, what are meaningful 
transformations of these random numbers to represent the emprirical information in an undistorted way? 
Hereby we understand a representation as undistorted, if the uncertainty volume of the representation 
vector\footnote{Note that the uncertainty volume of a random vector has nothing to do with the 
uncertainty relations of quantum theory.}, which is due to the finite empirical information, is constant 
for a given amount of empirical information and thus independent of the representation vector 
itself \cite{infoamount}.

We seek such a representation by making use only of the probabilistic paradigm 
of modern physics. We show that there is only {\em one} such way of representing observed data
and that the properties of the representation remain invariant only under linear transformations. In the 
limit of infinite empirical information this gives the state vector of quantum 
theory and its linear evolution. But interestingly, for finite information there should exist better 
representations. We comment on this in the discussion.

The paper is organized as follows: 

\noindent
{\em Section 2:} 

\noindent
Storing information from a probabilistic yes-no experiment. Encoding the relative frequency of 
yes and no into numbers with fixed credibility of the bits (or any other units).

\noindent
{\em Section 3:} 

\noindent
Vector respresentation of the empirical information. Easier to handle and more symmetry for
particular representations.

\noindent
{\em Section 4:} 

\noindent
Extension to probabilistic experiments with more than two outcomes. Generalisation of the
method of representation is straightforward, because the rule for encoding turns out to be the same
as for the yes-no experiment.

\noindent
{\em Section 5:} 

\noindent
Transformations of the representation vector. Linear transformations are preferable because
they introduce no unwanted structure in the representation of information. 

\noindent
{\em Section 6:} 

\noindent
Discussion.

\section{Storing information from a probabilistic yes-no experiment}
Given a probabilistic experiment with two possible outcomes, '0' and '1' (e.g. a Stern-Gerlach experiment
on a spin 1/2 particle, but for the present purpose tossing a biased coin is just as good). The 
probability $p$ of outcome '1' in a single trial is unknown, but known to have a definite value because all
experimental conditions are well controlled. We do $N$ trials in which 
'1' is obtained $L$ times (and '0' $N-L$ times). However, there are only $S$ bits of storage
available, and $S$ is too small to store the observed relative frequency $\nu \equiv L/N$ 
accurately. How should we encode the experimental result into the $S$ bits, such that the 
probability that these $S$ bits are correct, becomes maximal?

First, we simply store the relative frequency $\nu$ itself. That is, we round it to $S$ bits. Let us denote
this rounded number by $[\nu]_S$. Now we know
that in infinitely many trials $\nu$ would approach $p$. We can therefore trust $[\nu]_S$ to be correct,
if the difference between $\nu$ and $p$ is less than the value of the $S+1^{st}$ bit. In other words,
if
\be
\left|\nu-p\right| < \frac{1}{2^{S+1}}.
\ee
The probability that an experiment with $N$ trials will yield such a $\nu$ shall be denoted by
$Prob([\nu]_S)$. It is a function of $S$, $p$ and $N$,
\be
Prob([\nu]_S) = \sum_{L,(|L/N - p|<2^{-(S+1)})} \frac{N!}{L!(N-L)!} p^L (1-p)^{(N-L)}
\ee
where the summation is taken over those $L$ for which the condition is true. Fig.1, curve (a), 
shows this
probability as a function of $p$ for $N=4000$ trials and $S=6$ bits. (Exact storage
of a result would require log$_2$(4000)$\approx$12 bits.) Note that it is pretty low
around $p=0.5$, where it reaches only 0.68. This is because the fluctuation of the relative 
frequency $\nu$ is larger
for values of $p$ around $0.5$ than it is for values close to 0 or close to 1.

As a second example, we store the experimental result as quantum theory would suggest it. We
encode the observed probability {\em amplitude}. That is, we take $\eta \equiv \sqrt{\nu} = \sqrt{L/N}$
and round it to $S$ bits. The resulting number shall be denoted by $[\eta]_S$.
What is the probability that these $S$ bits are correct?

Here we must consider that in the limit of infinitely many trials the random number $\eta$ will approach
the limit $\sqrt{p}$. We can therefore trust $[\eta]_S$ to be correct,
if the difference between $\eta$ and $\sqrt{p}$ is less than the value of the $S+1^{st}$ bit. 
The probability that an experiment with $N$ trials will yield such an $\eta$ shall be denoted by
$Prob([\eta]_S)$. It is given by
\be
Prob([\eta]_S) = \sum_{L,(|\sqrt{L/N} - \sqrt{p}|<2^{-(S+1)})} \frac{N!}{L!(N-L)!} p^L (1-p)^{(N-L)}
\ee
where the summation is taken over those $L$ for which the condition is true. This probability is
shown in Fig.1, curve (b), again for $N=4000$ trials. Note that it is not symmetric about $p=0.5$. 
Its lowest value is for $p$ close to 0, where it drops to 0.65. This is lower than the lowest 
probability of 6 correct bits when storing the relative frequency directly.

\begin{figure}[ht]
\begin{center}
\epsffile{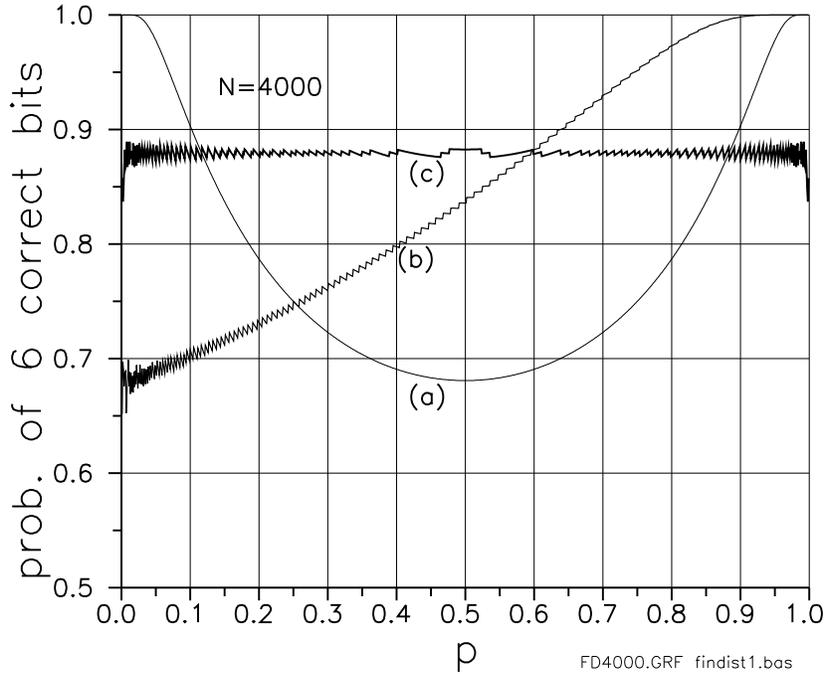}
\caption[Fig.1:]{ 
Probability of getting the first 6 bits correctly when taking a specific function of the
experimentally obtained relative frequency $L/N$ and decomposing it into binary form. Shown
as a function of the probability $p$. For $N=4000$ trials.
{\bf (a):} $L/N$. {\bf (b):} $\sqrt{L/N}$.
{\bf (c):} $\pi^{-1} \arcsin{(2L/N-1)} + 1/2$.
}
\end{center}
\end{figure}

As a third example we want to find that way of storing the experimental result, which can 
guarantee the highest minimum value of the probability that its first $S$ bits are correct.
We must find a smooth and monotonic mapping $\nu \rightarrow \chi$, where $\chi$ is also confined to
the interval [0,1], such that the largest fluctuations of the random variable $\chi$, that
can occur for any value of $p$, are smaller than for any other smooth and monotonic function
of $\nu$ in [0,1].

We argue as follows: The standard deviation of the relative frequency $\nu$ is well known as
\be
\sigma_{\nu} =\sqrt{\frac{p(1-p)}{N}}.
\ee
It is largest at $p=0.5$, which corresponds to a large fluctuation of the observed random 
variable $\nu$. In order to get a less fluctuating random variable $\chi$, it is therefore reasonable
to compress the region around $\nu \approx 0.5$ to a narrower region, and to expand the regions
close to 0 and close to 1. The compression-expansion factor should be proportional to $1/\sigma_{\nu}$.
Ideally, this should yield a random variable $\chi(\nu)$, whose fluctuations are independent of $p$.
The ratio of the standard deviations of $\chi$ and of $\nu$ should therefore be
\be
\frac{\sigma_{\chi}}{\sigma_{\nu}} = \frac{c}{\sqrt{p(1-p)}},
\ee
where $c$ is a constant.
In the limit of large $N$ this can be shown to yield the function
\be
\chi = \frac{1}{\pi} \arcsin{(2\nu - 1)} + \frac{1}{2} = \frac{1}{\pi} \arcsin{(2L/N - 1)} + \frac{1}{2}.
\ee
Fig.1, curve (c), shows the probability that the first 6 bits of this random variable are obtained correctly
in an experiment of $N=4000$ trials. Note that it is pretty constant at about 0.88 over the whole range
of $p$. The smallest values are approached close to 0 and close to 1, where it drops to 0.84. 

Clearly, $\chi$ is the best of the three investigated possibilities of storing the experimental result
of a probabilistic experiment when fewer storage bits are available than would be needed to encode
the experimental result precisely. And it seems that it is also the best conceivable way, because
the probability of getting the first 6 bits correctly tends to be constant. Any other function of
$\nu$ might improve this probability in some region of $p$, but necessarily at the expense of 
lowering it in another region of $p$.

Nevertheless, it is important to emphasize that the arcsine-relation of eq.(6), or its inverse,
\be
\nu = \sin^2(\frac{\pi}{2}\chi),
\ee
is only 
really the best function in the limit of infinitely many trials. But real experiments are always
finite. For these there exists an optimal function, whose form depends on the number of trials. It differs
from the sinusoidal relation for values of $p$ close to 0 and close to 1, where it is less curved.
(This will be the topic of a future paper \cite{QMcorrection}.)
 
It is also interesting to consider the conceptual status of the limit a random variable tends to in 
infinitely many trials. For the relative frequency $\nu$ this is the probability $p$. For the random 
variable $\chi$ it is a quantitity which we shall denote by $x$. The functional relation between the two is,
in analogy to the corresponding random variables, $p = \sin^2(\frac{\pi}{2} x)$. This is reminescent of
the quantum theoretical phase. But we should be cautious here. The quantity $x$ can be thought to exist
for $\em any$ probabilistic yes-no experiment, classical or quantum mechanical. It is simply the limit a
particular random variable tends to. However, it has a property, which no other limit of a
random variable possesses: The accuracy, with which it can be known, is knowable {\em before} the experiment
is done, because it is {\em invariant} of the probability $p$. (At least for infinitely many trials, but
it is a pretty good statement even for finitely many trials, as can be seen in the relative constancy of 
curve (c) in Fig.1).

\section{Vector representation }
Now the data of the yes-no experiment shall be represented as a two-component real vector. This is actually
an inefficient method, because the result of a yes-no experiment is only one random variable, not two. But
quantum theory suggests we should pay a closer look at such vectors. Clearly, though, the endpoint of such
a vector can only be along a line, not within an area.

In accordance with the previous section, we shall investigate the following three random vectors:
\be
\vec{\nu} \equiv \left( \begin{array}{c} \nu_1 \\ \nu_2 \end{array} \right) 
= \left( \begin{array}{c} \frac{L}{N} \\ 1-\frac{L}{N} \end{array} \right),
\ee
\be
\vec{\eta} \equiv \left( \begin{array}{c} \eta_1 \\ \eta_2 \end{array} \right) 
= \left( \begin{array}{c} \sqrt{\frac{L}{N}} \\ \sqrt{1-\frac{L}{N}} \end{array} \right)
\ee
and
\be
\vec{\chi} \equiv \left( \begin{array}{c} \chi_1 \\ \chi_2 \end{array} \right) 
= \left( \begin{array}{c} \frac{1}{\pi}\arcsin(2\frac{L}{N}-1)+\frac{1}{2} \\ 
\frac{1}{\pi}\arcsin(1-2\frac{L}{N}) +\frac{1}{2} 
\end{array} \right).
\ee
Here, $\vec{\nu}$ is the vector of relative frequencies of the two possible outcomes, $\vec{\eta}$ is the 
vector of the corresponding square roots (thus the probability amplitude representation of quantum theory,
except for phases), 
and $\vec{\chi}$ is the vector derived from our 'best' random variable of the previous section. Fig.2 shows
the lines of the possible end points of these vectors in the first quadrant of the real plane.

\begin{figure}[ht]
\begin{center}
\epsffile{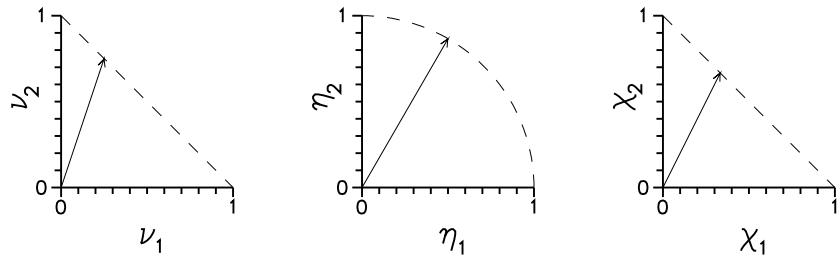}
\caption[Fig.2:]{ 
Graphical representation of the different random vectors $\vec{\nu}$, $\vec{\eta}$ and $\vec{\chi}$.
In all three graphs the vector corresponds to the
experimental result $L/N =.25$. {\small [vecnu.grf, veceta.grf, vecchi.grf]}
}
\end{center}
\end{figure}

We note that the end point of $\vec{\nu}$ can lie on a straight line of length $\sqrt{2}$. The same holds
for $\vec{\chi}$. And the endpoint of $\vec{\eta}$ can lie on a quarter circle of length $\frac{\pi}{2}$.

We pose the following question: What is the probability that the experiment of $N$ trials yields a vector
whose endpoint is no farther from the end point of the true vector than the fraction $2^{-(S+1)}$ of the
length of the line on which it can possibly lie? In other words, we ask, what is the probability that we
know the whereabouts of the true end point to an accuracy of $S$ bits after the experiment?

The answer for the vectors $\vec{\nu}$ and $\vec{\chi}$ can be given right away. It is the same as that
for the scalar quantities $\nu$ and $\chi$ of the previous section, because in each case we are just 
projecting the horizontal axis of the corresponding plot in Fig.2 to the line of possible end points. Since
these lines are straight, both for $\vec{\nu}$ and for $\vec{\chi}$, the statistical properties of scalar 
$\nu$, respectively $\chi$, are not distorted when going to vector $\vec{\nu}$, respectively $\vec{\chi}$. 
This means in particular that, 
in analogy to scalar $\chi$, for the vector $\vec{\chi}$ the probability that an experiment
will yield the whereabouts of its end point correctly to $S$ bits becomes an {\em invariant} of $p$ as
$N$ becomes large. This is evident in Fig.3, where this probability is shown as a function of $p$. (Note that
this probability is really the same as that for the scalar random variable $\chi$ in curve (c), Fig.1.)

The answer for the vector $\vec{\eta}$ must be sought more formally. We want to find the probabilty that
we can trust the experimentally found $\vec{\eta}$ to $S$ bits. This means we want to know the probability
for outcomes $L/N$, given $p$, for which
\be
\left| \left( \begin{array}{c} \sqrt{L/N} \\ \sqrt{1-L/N} \end{array} \right) -
\left( \begin{array}{c} \sqrt{p} \\ \sqrt{1-p} \end{array} \right) \right| < \frac{\pi}{2} 2^{-(S+1)},
\ee
where the factor $\pi/2$ is due to the fact that the endpoint of $\vec{\eta}$ is not confined to a curve
of length 1, but to one of length $\pi/2$. This probability shall be denoted by $Prob([\vec{\eta}]_S)$.
It is given by
\be
Prob([\vec{\eta}]_S) = \sum_{L_{(selected)}} \frac{N!}{L!(N-L)!} p^L (1-p)^{(N-L)},
\ee
where the summation is to be taken over those selected $L$ which fulfil condition (11). This probability
is also shown in Fig.3. 
\begin{figure}[h]
\begin{center}
\epsffile{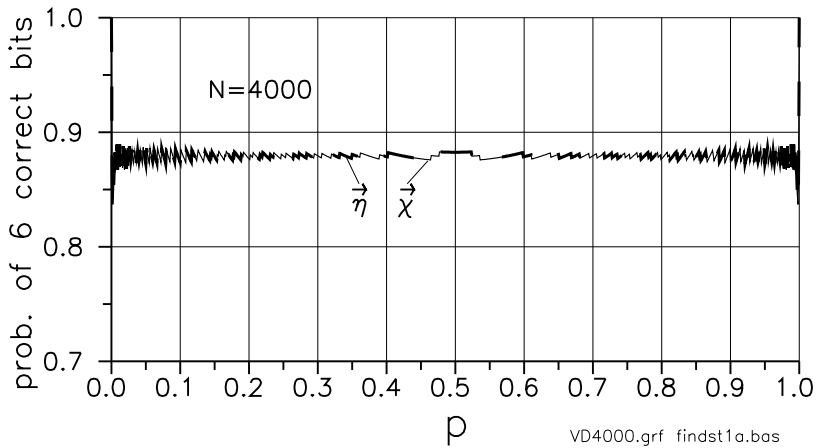}
\caption[Fig.3:]{ 
Probability of getting the position of the endpoints of the vectors $\vec{\eta}$ (thick dashed line) and 
$\vec{\chi}$ (thin line) correctly
to the first 6 bits in their respective domain from an experiment of $N=4000$ trials. Shown
as a function of the probability $p$. The curves coincide with each other.
}
\end{center}
\end{figure}
And note that it is exactly the same as that for $\vec{\chi}$. This means that in 
terms of accuracy of representation the vectors $\vec{\eta}$ and $\vec{\chi}$ are 
statistically equivalent
representations of the obtained information. The probability that the end point
of the vectors $\vec{\eta}$ and $\vec{\chi}$ will differ from the respective true end point (the
one approached in the limit of infinite trials) by less than a certain fraction of its possible range
becomes invariant of $p$ when $N$ is large. Then it depends only on $N$ and increases when
$N$ increases. That is why this 'confidence' probability can be specified without knowing the experimental
data. Knowledge of $N$ is sufficient. But the vector $\vec{\nu}$ does not have this invariance property.

A graphical way of understanding the statistical equivalence of $\vec{\eta}$ and $\vec{\chi}$ is to look how
$\vec{\chi}$ can be obtained from $\vec{\eta}$. One must only take the quarter circle on which the endpoint
of $\vec{\eta}$ lies, straighten it, and squeeze the resulting line from length $\frac{\pi}{2}$ homogeneously 
to length $\sqrt{2}$. This gives the line on which the endpoint of $\vec{\chi}$ lies.

But $\vec{\eta}$ has one additional feature of invariance, which $\vec{\chi}$ does not have. The length
of vector $\vec{\eta}$ is independent of the data. Interestingly, quantum theory seems to employ just this
vector (neglecting a complex phase factor), which not only represents the obtained information more 
accurately than virtually all others over the whole range of possible results, but which has one more 
symmetry over equivalent representations.

\section{Extension to a probabilistic experiment with K outcomes}
We shall now look at a general probabilistic experiment in which a single trial can give one out of $K$
possible outcomes. An example would be a projective measurement on a quantum-mechanical 
$K$-level system. (Note that even the most generalized modes of measurement are ultimately projective in
a higher dimensional Hilbert space than that of the original system.)
The probabilities for the $K$ different outcomes, $p_1,...,p_K$, whose sum is 1, are fixed by the 
preparation and the kind of projection done on the system. But they are unknown. 

In view of the specific invariance properties found for the vector $\vec{\eta}$ in the previous section, 
we shall here only investigate the multi dimensional extension of this representation vector. And in order
to be of general relevance to quantum theory, we add arbitrary complex phase factors to the components. Thus
$\vec{\eta}$ is now defined as
\be
\vec{\eta} = \left( \begin{array}{c} \sqrt{\frac{L_1}{N}}e^{i\varphi_1} \\ . \\ . \\ . \\ 
\sqrt{\frac{L_K}{N}}e^{\varphi_K} \end{array}
\right),
\ee
where $L_j$ denotes how often the outcome $j$ occurred in the $N$ trials, and the phases $\varphi_j$ are 
simply added on and cannot be determined in the projective measurement whose result $\vec{\eta}$ is to
represent. 

For reasons of analytical
simplicity we will shift our focus onto the {\em dispersion} of $\vec{\eta}$. We have already
remarked that when an experimentally obtained random number or random vector may have higher or
lower probability to be correct to a desired accuracy, this is a consequence of differing sensitivity of
the numerical decomposition to statistical fluctuations. Formally, these fluctuations are described by the 
dispersion or by its square root, the standard deviation. The reason why we found that the probability 
that the observed two component vectors
$\vec{\eta}$ and $\vec{\chi}$ differ by no more than $2^{-(S+1)}$ of their respective range from their
respective true vector becomes invariant of $p$, is that the dispersion for both $\vec{\eta}$ and 
$\vec{\chi}$ becomes invariant of $p$. And this is not only true for the two-component vectors, but also
for the corresponding vectors of arbitrary dimension, and even when we add arbitrary complex phases. We 
shall show this for the vector $\vec{\eta}$ of general dimension $K$.

First we must look at its expectation vector $E(\vec{\eta})$. (Whether the expectation $E(.)$ is a vector
or a scalar is determined by its argument.) The expectation value of the component 
$\eta_j \equiv \sqrt{\frac{L_j}{N}}e^{i\varphi_j}$ is defined as 
\be
E(\eta_j) = \sum_{L_1=0}^N ... \sum_{L_K=0}^N \frac{N!}{L_1! ... L_K!}
p_1^{L_1}...p_K^{L_K} \eta_j.
\ee
The multiple summation is subject to the constraint $\sum L_j = N$. It
can be greatly simplified by realizing that only the summation over $L_j$ takes
into account the factor $\eta_j$. Therefore, all other summations can be done
independently. This reduces the calculation of $E(\eta_j)$ to the case as if we were
doing an experiment with only two instead of $K$ outcomes. We only ask in each trial: Has
the outcome $j$ happened, yes or no? The statistics of this experiment is governed by the
binomial distribution, and so we can write, replacing the summation index $L_j$ by $l$, for simplicity,
\be
E(\eta_j) = \sum_{l=0}^N \frac{N!}{l! (N - l)!} (1-p_j)^{N-l}p_j^{l}\sqrt{\frac{l}{N}}e^{i\varphi_j}.
\ee
The calculation must be done numerically. We
emphasize that $E(\eta_j)$ is not identical to $\sqrt{p_j}e^{i\varphi_j}$ for small $N$, but approaches 
it for large $N$.

Now we turn to the dispersion of $\vec{\eta}$. It shall be denoted by $D^2(\vec{\eta})$. It is defined as the
expectation value of the quadratic difference between $\vec{\eta}$ and of the expectation of $\vec{\eta}$:
\be
D^2(\vec{\eta}) = E\left(\left|\vec{\eta} - 
E\left(\vec{\eta}\right)\right|^2\right).
\ee
Note that $\left|\vec{\eta} - E\left(\vec{\eta}\right)\right|^2$ is a 
real random number given by
\be
\left|\vec{\eta} - E\left(\vec{\eta}\right)\right|^2 = 
\sum_{j=1}^K \left|\eta_j - E(\eta_j)\right|^2.
\ee
Since the expectation value of a sum is equal to the sum of the expectation values we have
\be
E\left(\left|\vec{\eta} - 
E\left(\vec{\eta}\right)\right|^2\right) =
\sum_{j=1}^K  E\left(\left|\eta_j - E(\eta_j)\right|^2\right).
\ee
So we must only look at the formal calculation of the expectation value of the squared 
difference for {\em one} component of the vector $\vec{\eta}$. We label it 
$D_j^2$ and it is
\begin{eqnarray}
D_j^2 & = & E\left(\left|\eta_j - E(\eta_j)\right|^2\right) \nonumber 
\\ &=& 
\sum_{l=0}^N \frac{N!}{l! (N - l)!} (1-p_j)^{N-l}p_j^{l}\left|\eta_j -
E(\eta_j)\right|^2  \nonumber \\
&=& \sum_{l=0}^N \frac{N!}{l! (N - l)!} (1-p_j)^{N-l}p_j^{l}
\left[\frac{l}{N} -2Re\left(\eta_j^* E(\eta_j)\right)
+\left|E(\eta_j)\right|^2 \right],
\end{eqnarray}
where $\eta_j = \sqrt{l/N}e^{i\varphi_j}$.
The result is obtained numerically and is shown in Fig.4 as a function of $p_j$. We note
that, when multiplied by $N$, it approaches $\frac{1}{4}(1-p_j)$. And it is independent 
of the phase $\varphi_j$. With (18)
the dispersion of the {\em whole} vector $\vec{\eta}$, when also multiplied by $N$, therefore tends to a 
constant value, which is $\frac{K-1}{4}$. A deviation
exists only when one or several of the $p_j$ are close to 0, but it disappears when 
$N$ becomes large. We can therefore conclude, that the dispersion of the representation vector 
$\vec{\eta}$ of the result of a probabilistic experiment with $K$ different outcomes in a single 
trial tends to $\frac{K-1}{4N}$ when $N$ becomes large. It therefore tends to become an
{\em invariant} of the $p_j$. This means that the accuracy, with which the true vector is known
(i.e. the one which $\vec{\eta}$ approaches in the limit of infinitely many trials) only depends
on the number of trials. In other words, it is sufficient that we know $N$, in order to be able to
specify a small hypersphere around the endpoint of the experimentally determined
vector $\vec{\eta}$ within which the endpoint of the true vector will lie with a certain confidence
probability. As this confidence probability is independent of the $p_j$, for large $N$, it is also the 
highest achievable for any representation. So quantum theory picked a good way of representing empirical
information, indeed. (Having done this analysis I encountered a very illuminating paper by Caves and Fuchs 
\cite{FuchsCaves},
who defined the representation of the state vector by a finite number of bits as the {\it quantum information
content of the state}. In our case this would be the total number of bits with which we know $\vec{\eta}$ to
a certain confidence level, which are $(K-1)S$ bits, because we know no phases, and the $K^{th}$ component 
follows from unitarity.)
\begin{figure}[h]
\begin{center}
\epsffile{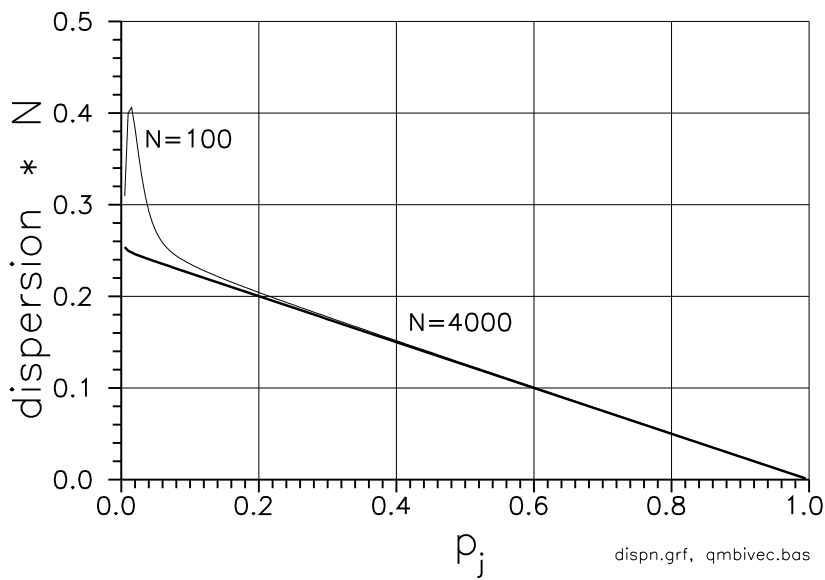}
\caption[Fig.4:]{ 
Dispersion $D_j^2$ of component $\eta_j \equiv \sqrt{\L_j/N}$, multiplied by number of trials,
as a function of $p_j$. Thin line: $N=100$. Thick line: $N=4000$.
}
\end{center}
\end{figure}

\section{Transformations of the representation vector}
Here we want to investigate which transformations can be made on $\vec{\eta}$ in order to obtain 
another vector $\vec{\psi}$ which has the same invariance properties as $\vec{\eta}$ and perhaps
even additional ones, and yet represents the empirical information without any loss. This means, once
$\vec{\psi}$ is obtained from the relative frequencies $\frac{L_1}{N},...,\frac{L_K}{N}$, it must be 
possible to get back out these relative frequencies when one is given only $\vec{\psi}$ and the arbitrary 
phases put into $\vec{\eta}$.

First, we will look at transformations for $K=2$. So we are again considering a yes-no experiment like
a projective measurement on a quantum mechanical 2-level system. Specifically, we consider the following
situation. A two-level system has been repeatedly prepared in some manner and each time we have done a 
certain measurement on it and so have obtained the vector $\vec{\eta}$. Now we want to do the whole 
experiment again, but instead of doing the
same projective measurement we let the system evolve for some time and then do this measurement. Does our
knowledge of $\vec{\eta}$ permit us to make any general statement of how the representation vector of the
result of the second measurement will look like? In other words, we are asking, whether we can find any
general rule of how the system will evolve in time, or to be even more precise, what our representation of
our knowledge of the system will look like as a function of the parameter time.

Well, a general rule can only be found if we adopt a general principle. And here it seems obvious to assume
that our knowledge of the sytem must not deteriorate in time. For, if the second measurement revealed that 
it did deteriorate, we would be forced to postulate that something unaccounted for has happened. 
In practice this means we would be forced to acknowledge that we were not aware of all the conditions the 
system was exposed to during the time interval of interest. Therefore,
we want to look for a transformation of the vector $\vec{\eta}$ into a vector $\vec{\psi}$, such that the
dispersion of $\vec{\psi}$ is the same as that of $\vec{\eta}$, and it must have the same invariance
property (i.e. it must not depend on the $p_j$, at least for large $N$).

Does the quantum mechanical rule of linear transformations conform to this principle? Here, a transformation
of $\vec{\eta}$ is effected by a general rotation
\be
{\bf R} = e^{i\vec{\sigma}\vec{\tau}}
\ee
where $\vec{\sigma}$ are the Pauli matrices, and $\vec{\tau}$ contains the duration, strength and direction
of the interaction. Writing out {\bf R} explicitly we have
\be
{\bf R} = \left( \begin{array}{cc}
\cos{\tau} + i\sin{\tau}\cos{\theta} & \sin{\tau}\sin{\theta}e^{-i\phi} \\
-\sin{\tau}\sin{\theta}e^{i\phi} & \cos{\tau} - i\sin{\tau}\cos{\theta}
\end{array}
\right),
\ee
where $\theta$ and $\phi$ specify the direction of $\vec{\tau}$ in polar coordinates and the scalar $\tau$
expresses the angle of rotation. Of course, we could 
have a succession of such rotations with different $\vec{\tau}$. The vector $\vec{\psi}$ now is
\be
\left( \begin{array}{c} \psi_1 \\ \psi_2 \end{array} \right) =
{\bf R} \left( \begin{array}{c} \eta_1 \\ \eta_2 \end{array} \right) =
{\bf R} \left( \begin{array}{c} \sqrt{\frac{L}{N}} e^{i\varphi_1} \\ \sqrt{1-\frac{L}{N}} e^{i\varphi_2}
\end{array} \right).
\ee
Its dispersion can be calculated in complete analogy to that of $\vec{\eta}$ (eqs.(14)-(19)). The result for
a specific rotation was
obtained numerically and is shown in Fig.5. The left drawing shows the dispersions of $\eta_1$, $\eta_2$ and
of the vector $\vec{\eta}$ itself. And the right drawing shows the corresponding dispersions for 
$\vec{\psi}$. Note that the dispersions for $\psi_1$ and $\psi_2$ show a different behaviour as a function
of the probability $p_1$ to obtain '1' in a single trial (of the first measurement!). But the dispersion of
the whole vector $\vec{\psi}$ tends to become an invariant of $p_1$ as the number of trials becomes large,
just like that of $\vec{\eta}$, and it also approaches the same value $\frac{1}{4N}$. Therefore, the
quantum mechanical evolution, at least for the two-level system, does ideed conform to the principle we 
hoped to see fulfilled, namely, that the input information is conserved. 
\begin{figure}[ht]
\begin{center}
\epsffile{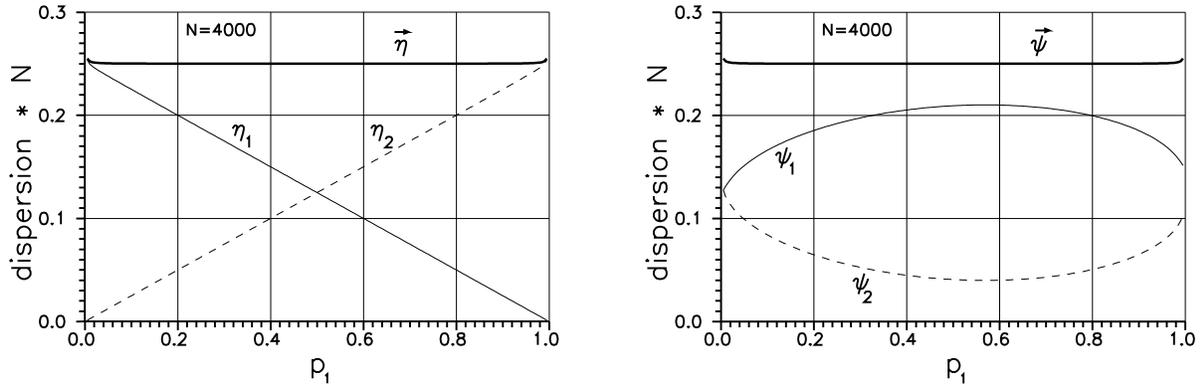}
\caption[Fig.5:]{Left side: Dispersions of $\eta_1$ (thin solid line), $\eta_2$ (thin dashed line) and
of $\vec{\eta}$ (thick horizontal line). Right side: Dispersions of $\psi_1$ (thin solid line), 
$\psi_2$ (thin dashed line) and of $\vec{\psi}$ (thick horizontal line). Both for $N=4000$ trials.
Parameters for rotation matrix {\bf R} in degrees: $\tau=75$, $\theta=50$, $\phi=110$.}
\end{center}
\end{figure}

The extension to the K-level system is straightforward. We can write any unitary transformation of the 
complex vector $\vec{\eta}$ with K components (eq.13) as a sequence of transformations applied to all
possible two-dimensional subspaces. Thus we have to define matrices $T_{ij}$, ($i=1,...,K-1$ 
and $j=i+1,...,K$), which are all equivalent to the K-dimensional
identity matrix, except that the elements $I_{ii}$, $I_{ij}$, $I_{ji}$ and $I_{jj}$ are replaced by the 
elements forming the 2-dimensional unitary matrix (eq.21), with suitably chosen parameters. There exist 
$K(K-1)/2$ such matrices $T_{ij}$. 
A method of constructing an arbitrary unitary $K \times K$ matrix as a product $\prod_{i,j}T_{ij}$ 
has been given by Reck et al. \cite{Reck}, following Murnaghan \cite{Murnaghan}.

It is now sufficient to realize that application of any of the $T_{ij}$ on an input vector $\vec{\eta}$
will result in a vector $\vec{\eta}'$, which is equivalent to $\vec{\eta}$ except for the $i^{th}$ and
$j^{th}$ components. 
In general, this will change the $i^{th}$ and $j^{th}$ components, and the dispersions of $\eta_i^{'}$ and
$\eta_j^{'}$ will not be the same as those of $\eta_i$, $\eta_j$, respectively, as can be seen in Fig.5. But
the sum of these dispersions does not change through the transformation, as we will show now. 
Following (17) and (18) the sum of the dispersions of $\eta_i$ and $\eta_j$ is
\be
D_i^2 + D_j^2 = E\left(\left|\eta_i - E(\eta_i)\right|^2\right) +
E\left(\left|\eta_j - E(\eta_j)\right|^2\right).
\ee
Abbreviating the general 2x2 rotation matrix, eq.(21) as
\be
\left( \begin{array}{cc}
a & b \\
-b^* & a^*
\end{array}
\right),
\ee
where we have $a^* a + b^* b = 1$, the transformed components are
$\eta_i^{'} = a \eta_i + b \eta_j$ and $\eta_j^{'} = -b^* \eta_i + a^* \eta_j$.
The sum of their dispersions is
\begin{eqnarray}
D_{i'}^2 + D_{j'}^2 & = & E\left(\left|\eta_i^{'} - E(\eta_i^{'})\right|^2\right) +
E\left(\left|\eta_j^{'} - E(\eta_j^{'})\right|^2\right) \nonumber \\
& = & E\left(\left|a\eta_i + b\eta_j - E(a\eta_i + b\eta_j)\right|^2\right) +
E\left(\left|-b^*\eta_i + a^*\eta_j - E(-b^*\eta_i + a^*\eta_j)\right|^2\right) \nonumber \\
& = & .... \nonumber \\
& = & E\left\{|\eta_i|^2 + |\eta_j|^2 + \left|E(\eta_i)\right|^2 + \left|E(\eta_j)\right|^2 -
2Re\left[ \eta_i E(\eta_i^*) + \eta_j E(\eta_j^*) \right] \right\}.
\end{eqnarray}
It is easy to see that this is the same as the sum of the dispersions of the original components,  eq.(23).
Therefore, the total dispersion of $\vec{\eta}'$ will be the same as that of $\vec{\eta}$. Thus the unitary 
transformation of a K-level system conserves the input information, as was the case for the 2-level 
system, above. This means the following: 
When we have done a projective measurement on $N$ identically prepared K-level systems and later prepare
copies in the same way, but let them evolve under well defined conditions before we do the projective 
measurement, our knowledge of the evolution law together with the input information obtained in the first
measurement enables us to specify the whereabouts of the true vector {\em after} the evolution with the same 
accuracy, as we were able to specify the true input vector. 

\section{Discussion}
We have set out with the conjecture that quantum theory is an optimal theory of encoding information obtained 
in the form of clicks, i.e. outcomes of probabilistic observations. For this purpose we have first looked 
for the most efficient way to represent data from a multinomial probability distribution by means of real 
(rational) numbers, because the statistics of quantum observations follows the multinomial distribution.
We asked how the observed relative frequencies should be mapped onto numbers, such that storing these
numbers by fewer bits than would actually be needed to store the relative frequencies exactly, ensures the
highest probability that these bits are correct (which means, that they coincide with those of the results
of an ideal experiment in which infinitely many trials can be done). We found that storing the vector whose
components are the square roots
of the relative frequencies is the most efficient way, provided the input data are obtained from sufficiently
many trials, because the statistical fluctuation of the endpoint of this vector, and thus the reliability of 
this information, becomes {\it invariant} of the probabilities behind the data. Next we investigated complex 
square roots of relative frequencies by adding arbitrary phase
factors. And instead of looking at the reliability of their bit-string representation we adopted the formal 
approach of looking at their dispersion. And here, too, we found that when representing the relative 
frequencies observed in a general probabilistic experiment by the vector of complex square roots of these 
relative frequencies, the dispersion of this random vector becomes {\it invariant} of the probabilities 
determining these relative frequencies. This is a very unique property, because it means that the accuracy 
of this representation of empirical information is independent of physical parameters.
It is interesting to note that quantum theory employs exactly
these vectors (or, to be exact, the limits they tend to in infinitely many trials), called {\it probability
amplitudes}, to describe a system.

We also investigated the properties of the random vector which results from a unitary transformation applied
to the vector of complex square roots of observed relative frequencies. It, too, showed the property that
its dispersion becomes an {\it invariant} both of the probabilities determining the input vector, as well as
of the parameters fixing the unitary transformation. Therefore, it would be an equally efficient way of
representing the empirical information. From the physical point of view this also means, that the quantum
mechanical evolution, which is described by just such a unitary transformation, preserves the information
we have about a system. If our original information is such that we can specify a small volume in Hilbert
space as containing the system, then the evolution will neither compress nor expand this volume, although
its shape may change. Note that this would in general not be true, {\em if we represented the system in any other
way than by the complex square roots of relative frequencies (probabilities)}.

Nevertheless, we also found that for observations with only few trials there should exist a better representation, 
which should deviate notably
from the complex-square-roots-of-relative-frequencies encoding, when these relative frequencies are due to 
probabilities close to 0 or close to 1. This could lead to an apparent deviation from the law of linear
superposition, e.g., when trying to predict the outcome of an experiment where a particle can fire 
a detector by reaching it over two different paths, or --- to extend it to entanglement --- when there exist two or 
more indistinguishable possibilities of how several particles can fire several detectors in coincidence). 
We may have measured the probabilities 
of each possibility separately, but with only few trials. And one of these probabilities may be very small. 
Then a simple adding of the complex square roots of the relative frequencies, even with suitable phases, 
may not be the most accurate prediction for the total probability amplitude. But one should not see this as a 
failure of quantum theory. It only tells us that quantum theory is a theory working with statistical limits.
Its statements refer to expectation values obtainable in infinitely many trials of probabilistic experiments.
We can rightly see it as the backbone of probabilistic science, because we can think that in principle any
individual observation can be repeated arbitrarily many times.
Only, if all our probabilistic observations were limited to only few trials, we cannot exclude
the possibility of a predictive theory which is more accurate than quantum theory.
We will look at this question elsewhere \cite{QMcorrection}.



\begin{thebibliography}{99}
\bibitem{infoamount} In this paper we mean by {\em amount of empirical information} the number of trials of
the probabilistic experiment from which the data were obtained. Thus the number $N$ or any monotonically 
increasing function of it. A more specific definition is not needed.
\end{thebibliography}
\end{document}